\documentclass[12pt,longbibliography,eprint]{revtex4-1}
\usepackage{epigraph}
\usepackage{graphicx}
\usepackage{epstopdf}
\usepackage{hyperref}
\usepackage{numprint}

\begin{document}
% $Id: SupplyDemand.tex,v 1.73 2016/02/14 06:36:58 mal Exp $
\preprint{V.M.}
\title{Market Dynamics. On Supply and Demand Concepts.}
\author{Vladislav Gennadievich \surname{Malyshkin}} 
\email{malyshki@ton.ioffe.ru}
\affiliation{Ioffe Institute, Politekhnicheskaya 26, St Petersburg, 194021, Russia}

\date{February, 14, 2016}

\begin{abstract}
\begin{verbatim}
$Id: SupplyDemand.tex,v 1.73 2016/02/14 06:36:58 mal Exp $
\end{verbatim}
The disbalance of Supply and Demand is typically considered
as the driving force of the markets.
However, the measurement or estimation of Supply and Demand
at price different from the execution price 
is not possible even after the transaction.
An approach in which Supply and Demand are always
matched, but the rate $I=dv/dt$ (number of units
traded per unit time) of their matching varies, is proposed.
The state of the system is determined not by a price $p$,
but by a probability distribution defined
as the square of a wavefunction $\psi(p)$.
The equilibrium state $\psi^{[H]}$
is postulated to be the one
giving maximal $I$ and obtained
from maximizing the matching rate functional $<I\psi^2(p)>/<\psi^2(p)>$,
i.e. solving the dynamic equation\cite{2015arXiv151005510G}
 of the form
``future price tend to the value
maximizing the number of shares traded per unit time''.
An application of the theory in a quasi--stationary
case is demonstrated.
This transition from
Supply and Demand concept to
Liquidity Deficit concept,  described by the matching rate $I$,
allows to operate only with observable variables,
and have a theory applicable to practical problems.
\end{abstract}

\keywords{Supply Demand, Liquidity Deficit, Market Dynamics}
\maketitle

\hfill\hbox{Dedicated to Nastya Tabakova}

\section{\label{intro}Introduction}
\begin{figure}
\includegraphics[width=8cm]{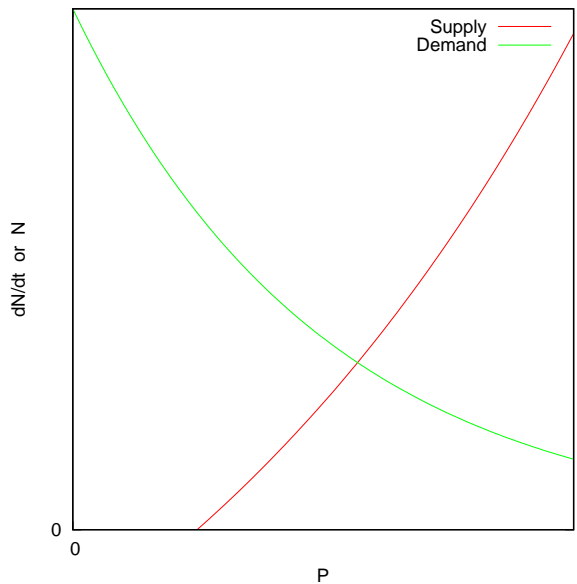}
\caption{\label{fig:qSD}
  Schematic plot of supply and demand as a function of price.
}
\end{figure}

The concept of Supply \& Demand is the central concept of modern economy.
With price increase the
production rate increases and consumption rate decreases.
The next step is to introduce the production rate (Supply curve $S(p)$)
and the consumption rate (Demand curve $D(p)$) as two functions of price,
see Fig. \ref{fig:qSD}, then consider their balance $S(p)=D(p)$ as
a stationary condition.
However, while the statement about production and consumption rate
is mostly correct, the introduction of supply $S(p)$ and demand $D(p)$ curves
poses severe limitation on a type of market dynamics and have been
criticized from a number of points.

Hans Albert \cite{albert2012model}, besides other problems,
point to the tautology
and interpretational problem with the approach,
so called ceteris paribus ("all other things being equal") problem,
that ``... theoreticians who interpret the clause differently de facto have different laws of
demand in mind, maybe even laws that are incompatible with each other.'' 
Joan Robinson  \cite{robinson1962economic}
point to a similar problem ``Utility is the quality in commodities that makes individuals want to buy them, and the fact that individuals want to buy commodities shows that they have utility''.
Another often discussed issues with classical
type of theory is equilibrium structure, supply--demand interdependence\cite{kirman1989intrinsic}
and adequacy to the real world markets\cite{goodwin2009microeconomics}.

We see the main problem with supply $S(p)$
and demand curves $D(p)$ concept that they are not
measurable or even observable at price different from current.
Even after the transaction is executed
we can tell nothing about $S(p)$ or $D(p)$
at price different from the execution price,
thus make the concept of ontological type, not applicable
to practical calculations.
The most intresting, the t\^{a}tonnement process\cite{walras2013elements},
as a mean to observe the supply/demand curves
misses the whole aspect
of market dynamics\cite{donier2015walras}.
In our initial approach\cite{2015arXiv151005510G} to build a
dynamic theory based on observable variables
the importance of execution rate $I=dv/dt$,
the number of entities (e.g. equity shares) traded
per unit time was emphasized, and the dynamic equation
of the form ``future price tend to the value
maximizing the number of shares traded per unit time''
was postulated and then, to some degree, observed experimentally.
That paper was dealing with complicated issues
of realtime HFT trading, so the calculations there
were performed in P\&L space, not in price space
(for a trader asset price is irrelevant, only the P\&L is important),
specific time--dependent basis mixed with trading signals was chosen.
In this paper we  consider
a simplified, quasi--stationary problem,
where the $I$ is assumed to be only a function of price $I(p)$,
then we  propose the characteristics,
estimating equilibrium properties
of the market.
As an example  US equity market for AAPL stock on
the September, 20, 2012, same day used in \cite{2015arXiv151005510G}
will be considered.
We understand this market is actually does not have true
equilibrium, but it allows us to demonstrate the technique
to calculate equilibrium state from the data
and show the behaviour of such characteristics as $I$, probability,
and price ``dynamic impact''.

\section{\label{trade}Trading rate in a quasi--stationary case}
As we postulated in \cite{2015arXiv151005510G}
the dynamic equation has the form:
\begin{eqnarray}
  I(p)&\to&\max \label{Ip} \\
  I&=&\frac{dv}{dt}
\end{eqnarray}
where $I$ is the number of units $dv$ traded
per unit time $dt$. The $I$ describe the rate of buyers and sellers matching
(a sell/buy market order matched a buy/sell limit order, already in the order book).
With this definition of $I$ the buyers and the sellers are always matched.
There is no any buyers--sellers disbalance at any price.
What is a function of price --- the rate $I$, at witch
buyers and sellers match each other.
Instead of using an assumption of buyers--sellers (demand--supply)
disbalance,  we consider  always matched buyers--sellers
and maximize the rate of matching, postulating that such a state of maximal $I$
to be the equilibrium state.
In a quasi--stationary case, considered in
this paper, the $I(p)$ is assumed to be a function of price only,
not depending on time explicitly (this assumption is incorrect in a general case).
Introduce a polynomial basis $Q_k(p)$ with $k=[0..d-1]$
(the results are invariant with respect to a linear basis transform,
so as a $Q_k(p)$ an arbitrary polynomial of $k$--th degree
can be chosen, e.g. $p^k$, but in practice
the basis selection is determined
by the numerical stability of calculations).
Two measures need to be used for this theory. In the simplest
form they both are a sum over all observation points,
but different with $dt$ and $dv$ type of integration:
\begin{eqnarray}
  <f>_{t}&=&\sum_{q} (t_q-t_{q-1})f(t_q)  \label{mut} \\
  <f>_{v}&=&\sum_{q} (v_q-v_{q-1})f(t_q) \label{muv}
\end{eqnarray}
The (\ref{mut}) integrate over $dt=(t_q-t_{q-1})$
and the (\ref{muv}) integrate over $dv=(v_q-v_{q-1})$,
index $q$ label observation event of matching
market order with limit order already in the order book.
Two Gramm matrices are defined on these two measures
\begin{eqnarray}
  G^{t}_{jk}&=&<Q_j Q_k>_{t} =
  \sum_{q} (t_q-t_{q-1})Q_j(p(t_q))Q_k(p(t_q))
  \label{Gt} \\
  G^{v}_{jk}&=&<Q_j Q_k>_{v} =
  \sum_{q} (v_q-v_{q-1})Q_j(p(t_q))Q_k(p(t_q))
  \label{Gv}
\end{eqnarray}
The key element of the theory is an introduction of
a wavefunction state $\psi(p)$,
determining probability density,
similar to the approach
of using quantum mechanics --like probability state to describe classical measurement
experiment\cite{malyshkin2015norm}
\begin{eqnarray}
  \psi(p)&=&\sum_{k=0}^{d-1}\psi_k Q_k(p)
  \label{psi}
\end{eqnarray}
Each probability state (\ref{psi}) determine the value of $I$
\begin{eqnarray}
  I_{\psi}&=&\frac{\left<\psi^2(p) I\right>_t}{\left<\psi^2(p)\right>_t}
  =\frac{\left<\psi^2(p) \right>_v}{\left<\psi^2(p)\right>_t}
  \label{Ipsi}
\end{eqnarray}
here the $\psi^2(p)$ can be treated as a probability density.
Using Gramm matrices definitions (\ref{Gt}) and (\ref{Gv})
the (\ref{Ipsi}) can be expressed as a ratio of two quadratic forms,
estimator of a stable form\cite{malha}:
\begin{eqnarray}
  I_{\psi}&=&\frac{\sum\limits_{j,k=0}^{d-1} \psi_j  G^{v}_{jk} \psi_k}
  {\sum\limits_{j,k=0}^{d-1} \psi_j  G^{t}_{jk} \psi_k}
  \label{IpsiG}
\end{eqnarray}
Given the (\ref{IpsiG}) mapping of a $\psi$ state to the value of $I$
the question arise about the $\psi$ states of maximal $I$ (corresponding
to (\ref{Ip}) dynamic equation), of minimal $I$ (corresponding
to liquidity deficit state) and of a given price value state.

The states, corresponding to minimal and maximal $I$
can be found from generalized eigenvalues problem
\begin{eqnarray}
  \sum\limits_{k=0}^{d-1}G^{v}_{jk} \psi^{[i]}_k&=&
    \lambda^{[i]} \sum\limits_{k=0}^{d-1} G^{t}_{jk} \psi^{[i]}_k
  \label{GevProblem}
\end{eqnarray}
Where the $(\lambda^{[i]},\psi^{[i]}(p)) ; i=[0..d-1]$ pairs
define the value of $I$ and corresponding to it probability
distribution $\left(\sum_{k=0}^{d-1}\psi^{[i]}_k Q_k(p)\right)^2$.

The $\psi^{[H]}(p)$ state, corresponding to maximal
$\lambda$ (the maximal $I$) is special,
it corresponds to the equilibrium state.
This is a replacement of classical supply--demand theory
where the equilibrium
is determined by price, obtained from $S(p)=D(p)$ equation.
In our new approach the equilibrium is defined
not by a specific price $p$,
but by the probability distribution $\left(\psi^{[H]}(p)\right)^2$ obtained
as the
eigenvector of (\ref{GevProblem}) problem.
The price, or any other observable variable,
corresponding to this state, can be
calculated in a similar to (\ref{Ipsi}) way, e.g.
\begin{eqnarray}
  p_{\psi^{[H]}}&=&\frac{\left<\left(\psi^{[H]}(p)\right)^2 p\right>_v}{\left<\left(\psi^{[H]}(p)\right)^2\right>_v}
  \label{PpsiH}
\end{eqnarray}
The value of any observable variable in equilibrium
can be calculated from (\ref{Ipsi}) by replacing the $I$ by the value of interest.

A typical application of (\ref{Ip}) dynamic equation
to a quasi--stationary
problem consists of calculating  from observation data
the  $G^{t}_{jk}$ and 
$G^{v}_{jk}$ matrices,
solving the Eq. (\ref{GevProblem}) problem,
obtaining the equilibrium state $\psi^{[H]}(p)$.
After the $\psi^{[H]}(p)$ is found all the
observable variables of interest can be calculated.
An important feature of all the $\psi^{[i]}(p)$ states, including the $\psi^{[H]}(p)$,
is that the first variation of $I$ from (\ref{Ipsi}) on these states is equal to zero
for arbitrary variation $\delta \psi(p)$,
what immediately follows from the (\ref{GevProblem}):
\begin{eqnarray}
  \delta \frac{<(\psi^{[i]}(p))^2>_v}{<(\psi^{[i]}(p))^2>_t}&=&
  2\left(\frac{<\psi^{[i]}(p) \delta \psi(p) >_v}{<(\psi^{[i]}(p))^2>_t} -
  \lambda^{[i]}
  \frac{<\psi^{[i]}(p) \delta \psi(p) >_t}{<(\psi^{[i]}(p))^2>_t}
  \right)=0
  \label{varI0}
\end{eqnarray}
The second variation of  $I$ from (\ref{Ipsi}) on $\psi^{[i]}(p)$ states
is:
\begin{eqnarray}
  \delta\delta \frac{<(\psi^{[i]}(p))^2>_v}{<(\psi^{[i]}(p))^2>_t}&=&
  2\left(\frac{<(\delta \psi(p))^2 >_v}{<(\psi^{[i]}(p))^2>_t} -
  \lambda^{[i]}
  \frac{<(\delta \psi(p))^2 >_t}{<(\psi^{[i]}(p))^2>_t}
  \right)
  \label{var2I0}
\end{eqnarray}
The other second variation terms vanish because of (\ref{GevProblem}).
For  for $i=H$ the (\ref{var2I0}) is always negaive.

The variation of $p_{\psi^{[H]}}$ from (\ref{PpsiH}) can be
calculated by considering the $\psi^{[H]}(p)+\delta \psi(p)$ states
(without losing a generality the $\delta \psi(p)$ here can be
considered orthogonal to $\psi^{[H]}(p)$, i.e.
$\delta \psi(p)=\sum\limits_{i=0; i\ne H}^{d-1}\beta^{[i]} \psi^{[i]}(p)$)
and the answer in the second order perturbation theory is:
\begin{eqnarray}
  p_{\psi^{[H]}+\delta \psi}&=&
  p_{\psi^{[H]}}+\frac{2}{\lambda^{[H]}}\left<\psi^{[H]}\left(pI\right)\delta \psi\right>_t
  +\frac{1}{\lambda^{[H]}}\left<\delta \psi\left(pI - p_{\psi^{[H]}} I\right)\delta \psi\right>_t   \label{varPpsiH} \\
  &=&  p_{\psi^{[H]}}+
  \frac{2}{\lambda^{[H]}}\sum\limits_{i=0; i\ne H}^{d-1}\beta^{[i]}\left(pI\right)^{[iH]}
  +\frac{1}{\lambda^{[H]}}\sum\limits_{i,l=0; i,l\ne H}^{d-1} \beta^{[i]}\left(pI - p_{\psi^{[H]}} I\right)^{[il]} \beta^{[l]}\nonumber\\
  \left(pI\right)^{[il]}&=&< \psi^{[i]}(p) p  \psi^{[l]}(p)>_v \label{ddm} \\
  \left(pI - p_{\psi^{[H]}} I\right)^{[il]}&=&< \psi^{[i]}(p) \left(p-p_{\psi^{[H]}}\right)  \psi^{[l]}(p)>_v \label{lsmatr}
\end{eqnarray}
The (\ref{varPpsiH}) is a quadratic function on $\beta^{[i]}$,
and the extremum of (\ref{varPpsiH}) can be found
by solving a linear system of $d-1$ size (the $\beta^{[H]}=0$,
what reduces the size of the system by 1), then substituting
the $\beta^{[i]}$ found back to (\ref{varPpsiH}) to find the value of the extremum of
$p_{\psi^{[H]}+\delta \psi}$.
\begin{eqnarray}
  &&Ex(p_{\psi^{[H]}+\delta \psi})=
  p_{\psi^{[H]}}-\frac{1}{\lambda^{[H]}}
  \sum\limits_{i,l=0; i,l\ne H}^{d-1} \left(pI\right)^{[Hi]}
  \left(\left(pI - p_{\psi^{[H]}} I\right)^{-1}\right)_{[il]} \left(pI\right)^{[lH]}
  \label{pe}
\end{eqnarray}
If linear system matrix (\ref{lsmatr})
is degenerated this means that price variation is rather small
and no information about price movement caused by 
execution flow spikes can be obtained.
If the extremum of $p_{\psi^{[H]}+\delta \psi}$
is equal to $p_{\psi^{[H]}}$ this means that all the (\ref{ddm}) elements
$\left(pI\right)^{[iH]}=0; i\ne H$  vanish,
what means that  $I$ and $p_{\psi^{[H]}}$ reach the extremum
on the same state $\psi^{[H]}$.

The question arise
of determining the state $\psi_{P}(p)$,
corresponding to the specific price value $P$.
In the simplest case
Radon--Nikodym type of approximation\cite{2015arXiv151005510G}
can be used:
\begin{eqnarray}
  \psi_{P}(p)&=&\frac{\sum\limits_{j,k=0}^{d-1} Q_j(p) \left(G^{t}\right)_{jk}^{-1}  Q_k(P)}
      {\sqrt{\sum\limits_{j,k=0}^{d-1} Q_j(P) \left(G^{t}\right)_{jk}^{-1}  Q_k(P)}}
  \label{psiP}
\end{eqnarray}
Substitution of (\ref{psiP}) to (\ref{Ipsi})
give an estimate of the number of units traded per unit time at price $P$:
\begin{eqnarray}
  I(P)&=&\frac{\sum\limits_{j,k,l,m=0}^{d-1} Q_j(P) \left(G^{t}\right)_{jk}^{-1}
    G^{v}_{kl} \left(G^{t}\right)_{lm}^{-1} Q_m(P)}
  {\sum\limits_{j,k=0}^{d-1} Q_j(P) \left(G^{t}\right)_{jk}^{-1}  Q_k(P)}
  \label{IpRN}
\end{eqnarray}

Another important application of (\ref{psiP})
is related to probability density analysis.
The $\left(\psi^{[i]}(p)\right)^2$
is unbounded
and is hard to scale. It is much more convenient for an analysis (and graphical representation)
to use a squared projection  of $\psi^{[i]}(p)$ on $\psi_{P}(p)$:
\begin{eqnarray}
  w^{[i]}(P)&=&<\psi^{[i]}(p)\psi_{P}(p)>_t^2 = \nonumber\\
  &&\frac{\left(\sum\limits_{k=0}^{d-1}\psi^{[i]}_k Q_k(P)\right)^2}
       {\sum\limits_{j,k=0}^{d-1} Q_j(P) \left(G^{t}\right)_{jk}^{-1}  Q_k(P)}=
       \frac{\left(\psi^{[i]}(P)\right)^2}
       {\sum\limits_{j,k=0}^{d-1} Q_j(P) \left(G^{t}\right)_{jk}^{-1}  Q_k(P)}
       \label{wP}
\end{eqnarray}
that is bounded to $[0..1]$ interval.
The $w^{[i]}(P)$ has a meaning of probability:
how close are the two states: the $\psi_{P}(p)$, the one with price equal to $P$,
and the $\psi^{[i]}(p)$,
the (\ref{GevProblem}) eigenstate.

\section{\label{numestim}Numerical Estimation of the Equilibrium}
As application example of this theory
let us apply it to AAPL stock data on September, 20, 2012.
This data is definitely not a  quasi--stationary,
but let us forget about this for a moment and use the data as an illustrative example
of the technique.
The calculations are performed in the following way.
Obtain every observation tick, labeled by $q$ index,
as a triple of time,price traded, total volume since the beginning
$(t_q,p_q,v_q)$.
The volume traded at $q$--th tick is $v_q-v_{q-1}$ and the time
passed between $q$--th and $(q-1)$--th ticks is $t_q-t_{q-1}$.
Having all this data available, choose a polynomial basis $Q_k(p)$
(the $Q_k(p)=p^k$ choice cause severe numerical instability for $d>3$,
 see  
 Appendix A of \cite{2015arXiv151005510G}
 with a few examples of stable basis selection),
and solve generalized eigenvalue problem (\ref{GevProblem})
using standard, e.g. LAPACK\cite{lapack} routines dsygv,  dsygvd and similar.
Among the eigenvectors of the (\ref{GevProblem})
select the $\psi^{[H]}(p)$ corresponding to maximal $\lambda$, that
give the equilibrium state probability
distribution. Then all the variables of interest
can be calculated from this $\psi^{[H]}(p)$.

% set output "q.eps" ; set terminal postscript eps size 12cm,6cm enhanced color ; set xrange [9.5:16]; set yrange [693:701] ; set xtics 0.5 ; set grid ;  set ytics 0.5 ; set ylabel "P" ;  set xlabel "t" ; l=694; s=0.5;cx=0.7; oo=0.7;ox=0.5; plot   "/tmp/k.dat" using ($2/3600e9):($5) with lines title "P"
\begin{figure}
\includegraphics[width=18cm]{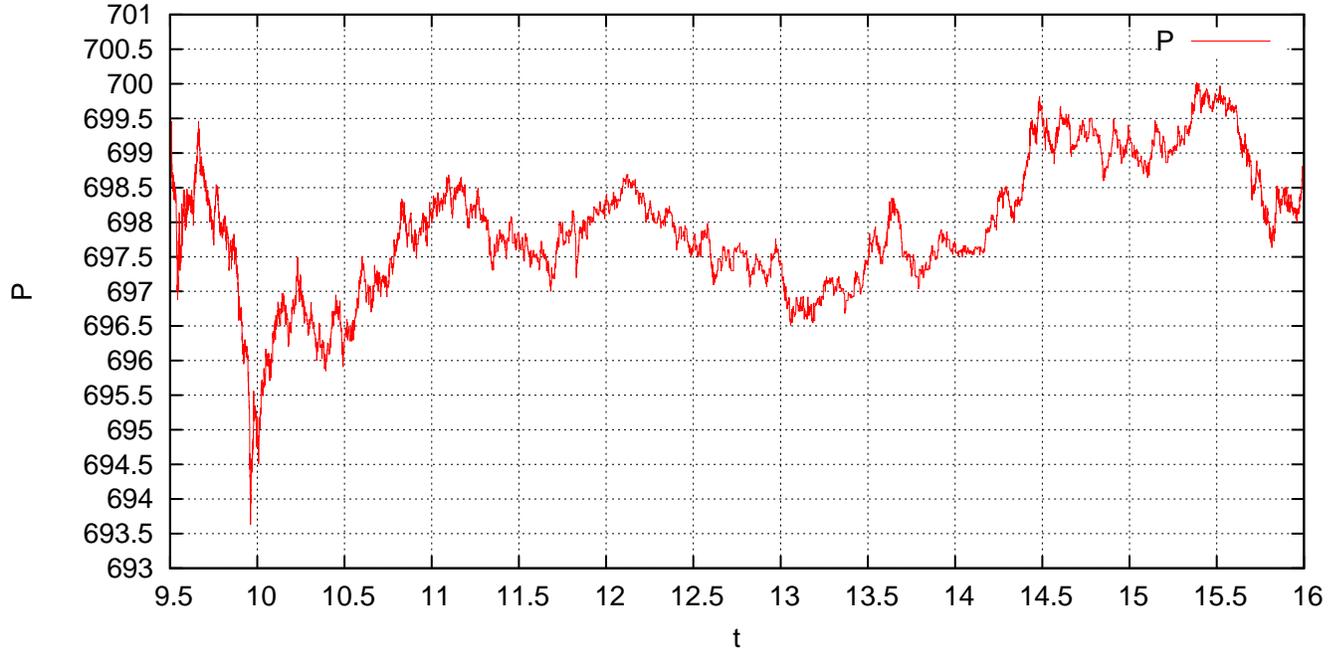}
\caption{\label{fig:P}
  The AAPL stock price on September, 20, 2012.
  The time on x axis is in decimal fraction of an hour, e.g. 9.75 mean 9:45am.
}
\end{figure}
In Fig. \ref{fig:P} the price of AAPL stock is presented as a function
of time. This day was specifically chosen to have
trending and volatility periods.

\begin{figure}
%set output "q.eps" ; set terminal postscript eps size 12cm,6cm enhanced color ; l=696; s=0.5;cx=0.7; oo=0.7;ox=0.5; set xrange [693:701] ; set xlabel "P" ;  plot   "histogram.csv" using (($1+$2)/2):($3)  with boxes title "Volume"
  \includegraphics[width=10cm]{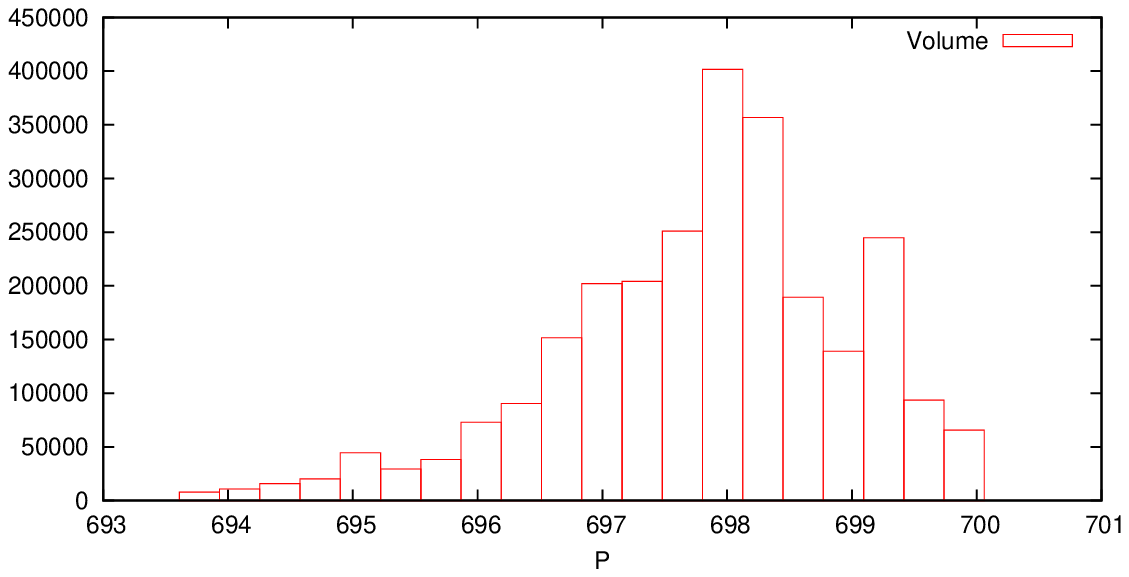}
%set output "q.eps" ; set terminal postscript eps size 12cm,6cm enhanced color ; l=696; s=0.5;cx=0.7; oo=0.7;ox=0.5; set xrange [693:701] ; set grid ;set xlabel "P" ; plot  "psi.csv" using ($1):($2)  with lines title "I(P)"
  \includegraphics[width=10cm]{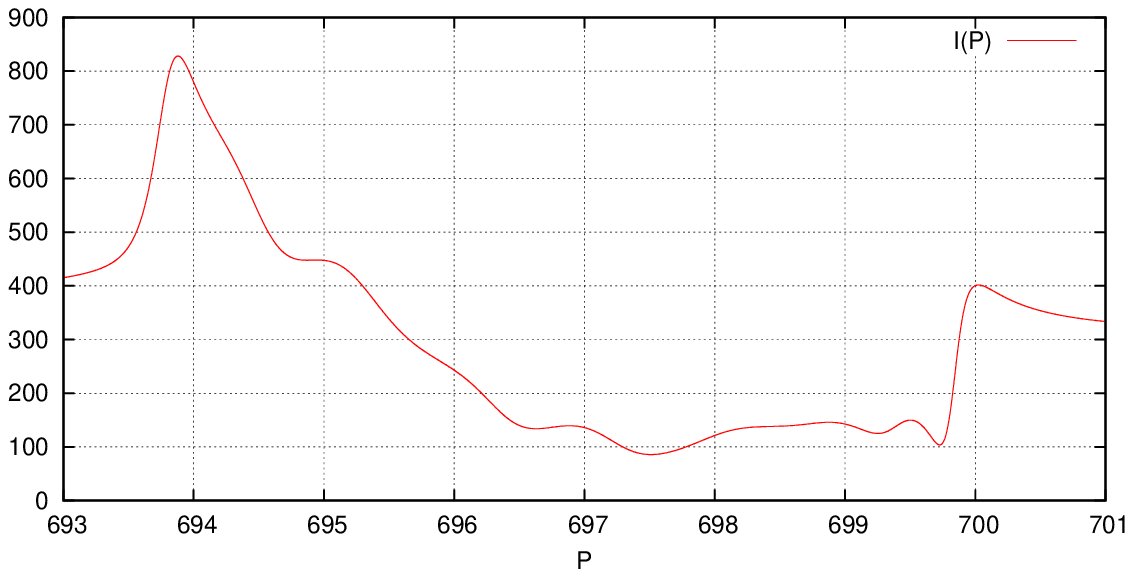}
%set output "q.eps" ; set terminal postscript eps size 12cm,6cm enhanced color ; l=696; s=0.5;cx=0.7; oo=0.7;ox=0.5; set yrange [0:1] ; set xrange [693:701] ; set grid ;set xlabel "P" ; plot  "psi.csv" using ($1):($6)  with lines title "w^{[H]}(P)" ,   "psi.csv" using ($1):($4)  with lines title "w^{[L]}(P)"
   \includegraphics[width=10cm]{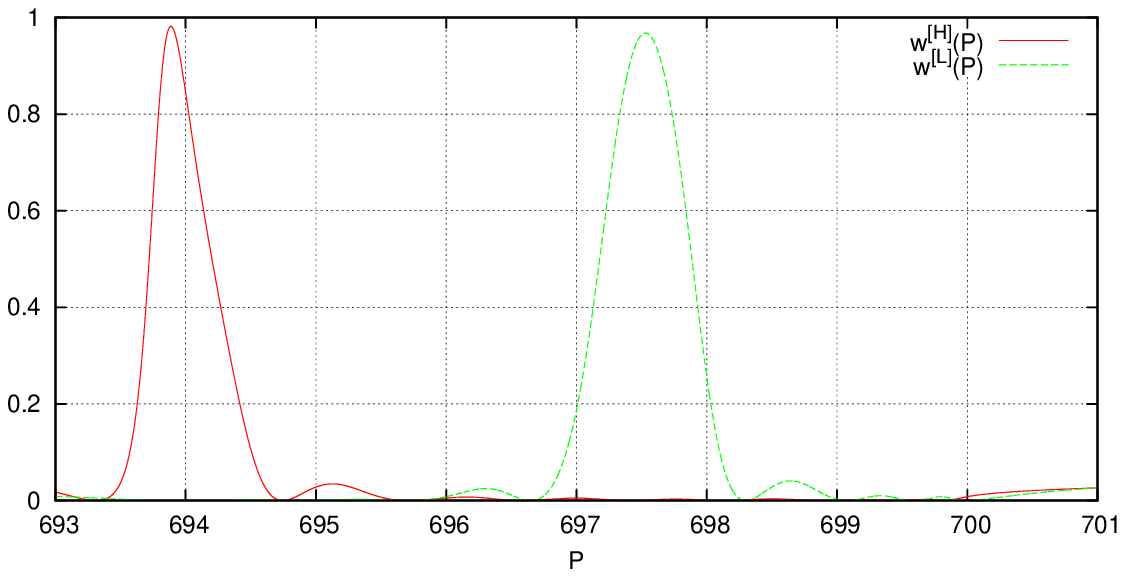}
\caption{\label{fig:EQl}
  Price histogram for volume distribution.
  $I(P)$ from (\ref{IpRN}) as a function of price.
  Projections (\ref{wP}) to the states corresponding to
  highest $w^{[H]}(P)$ and lowest $w^{[L]}(P)$ value of $I$.
}
\end{figure}

In top chart in Fig. \ref{fig:EQl}
a histogram of volume
distribution is presented. The sum of all histogram columns give total volume
(equal to 2,630,738 shares).
reported by NASDAQ ITCH feed between 9:30 and 16:00 on September, 20, 2012.
(Total AAPL  traded shares reported by ITCH feed on that
day, including off-market hours, is 3,063,928 ;
Google Finance (84,141,932) \& Yahoo Finance (84,142,100)
report higher volume, as consolidated from a number of exchanges).
Price analysis of various kinds, e.g. skewness, fat tails analysis, etc.
is often performed by analysts considering the charts
similar to this price--volume distribution.
We want to stress, that the volume distribution
carry no information about the dynamics.
First, this distribution assume some manually selected
time interval for the analysis. This make
the result observer--dependent.
Second, maximization of trading volume
is equivalent to ``buy below median price'' and
``sell above median price'' type of trading strategy.
This type of strategy assume the median price,
that is a non--local value (depend on the entire observation set), is known,
what is not the case for market dynamics, that
is determined by the events of a local nature on relatively small
(and unknown upfront) time scales.

In the low charts in Fig. \ref{fig:EQl} the results
for the  theory
are presented for $d=10$ elements in basis.
In the second chart in Fig. \ref{fig:EQl}
the $I(P)$ at given price $P$ is calculated
using the state (\ref{psiP}) to receive (\ref{IpRN}).
One can see a much sharper
picture, than the one with the volume histogram.
There are high values of $I$ near the 694 and 700 price levels.
These levels correspond to a very active trading (large $I$)
at around 10am and near market close at 15:30.
The theory automatically select the price levels of
high trading activity, not the high total volume traded
as in the chart above, that
provide volume median at
about the 698 level. Most of the shares were actually traded
near this, close to median, price.

There is no singularity in total volume distribution near 694 and 700 price levels,
because despite a high value of  $I$ the total time spent
at these price levels is low, what result in low volume traded.
But in terms of price change, risk analysis and market dynamics the volume per unit time,
not the total volume should be used, because
this characteristics describe market activity at given price level.
From these two charts (volume distribution and $I(P)$)
we can make an important observation on price impact.
The price impact \cite{2009PhRvE..80f6102M,gatheral2013dynamical,2014arXiv1412.0141D}
is typically considered as path--dependent impact
of executed shares on asset price.
 As we see from the volume distribution chart
the volume near 694 level is rather low, but
the price change and $I$ are high.
This make reasonable to introduce
a ``dynamic impact'' concept, the sensitivity
of asset price to the rate $I$,
different from regular ``impact'' ---
the sensitivity of asset price to executed volume
\cite{wiki:marketimpact}.
From the charts presented we see an importance
of the ``dynamic impact'' concept,
as both price and $I$ singularities are localized at about the
same price levels. Similar situation occur in the time--space,
\cite{2015arXiv151005510G} what make one more
argument, that price change is related to $I$, not to the volume
and the ``dynamic impact'' should be considered
as the major contribution of trading activity
affecting asset price.
The Eq. (\ref{pe}) give an opportunity
to experimentally answer on the fundamental question
whether the extremum of $I$ matches the extremum of $p_{\psi^{[H]}}$.
In classical impact \cite{wiki:marketimpact} model (increase in trading volume
causes substantial price changes) the extremums of $I$ and $p_{\psi^{[H]}}$
should be substantially different.
In the ``dynamic impact'' model (an increase in $I$ causes
substantial price changes) the extremums of $I$ and $p_{\psi^{[H]}}$
correspond the same state $\psi^{[H]}$.
For the data we used the values  $p_{\psi^{[H]}}=693.96$
and $Ex(p_{\psi^{[H]}})=692.46$ were obtained. This, along with
observation of large price movement near large $I$
make the ``dynamic impact'' model seems to be more appropriate.
However, the  wavefunction in price space are applicable
only to a quasistationary case and different basis
need to be considered when trying to make a
conclusion about a relative importance of
``classical impact'' and ``dynamic impact''
in a non--stationary case.

The transition from total volume (number of shares)
to volume traded per unit time allows to
overcome two mentioned above limitations of price--volume analysis. 
First, in contrast with price median,
that has no degree of freedom, and is determined
only by the manually inserted time scale, the (\ref{GevProblem})
eigenvalues problem have the $d$ degrees of freedom,
what allows automatic selection of  the most appropriate
time scale according to the dynamic equation (\ref{Ip}).
Second, in contrast with the total volume traded
the trading rate $I=dv/dt$ is local, a few large trade executed
in a short time period can drastically change
the value of $I$.
One can see this transition from volume to volume per unit time
as some kind similar
to the transition of momentum of motion concept
from Aristotle to Galilei type of classical mechanics.

In the third chart in Fig. \ref{fig:EQl} we present
the probabilities $w^{[H]}(P)$
and  $w^{[L]}(P)$ of a given price $P$
to correspond to the state of maximal and minimal $I$
respectively.
The  $w^{[H]}(P)$
is localized
in the area of high $I$ and large price changes
(the ``dynamic impact'' effect).
The $w^{[L]}(P)$ is localized in the area of low $I$,
in a quasi--stationary case the probability $w^{[L]}(P)$
has no importance, but it is
extremely important in a non--stationary
case\cite{2015arXiv151005510G}, as determining
trading position opening condition.
Market equilibrium interpretation is 
now have a probabilistic nature and
is determined by a wavefunction  $\psi^{[H]}(p)$
of the equilibrium state, corresponding to maximal $I$.
This is a major transition from price--space classical equilibrium
$S(P)=D(P)$ to the dynamic equation of (\ref{Ip}) form.
The goal of this transition
is to obtain the equilibrium state directly
from market data and, as we demonstrated above with the $\psi^{[H]}(p)$
calculation, this can performed by solving the
(\ref{GevProblem}) problem.
This is a critical step forward compared to 
the estimation of classical supply $S(P)$ and demand $D(P)$ functions
from market data, what is
ambiguous at best and impossible at worst.
The transition from supply/demand to matching rate
shift the study of $S(P)=D(P)$ in the vicinity
of equilibrium (e.g. whether it has linear or square root type of
behavior\cite{donier2015walras})
to the study of $I$ in the vicinity of $\psi^{[H]}(p)$,
because, as we demonstrated experimentally (see Appendix \ref{code} for code and data),
the dynamic changes in price due to $I$ changes are much greater
than price changes due to volume changes.
In this paper only quasi--stationary case is considered,
what lead to only two types of $I$ variations: Eq. (\ref{varI0}) and
(\ref{var2I0}), with $\delta \psi(p)$ as the $\psi^{[H]}(p)$ variation
in price space. For the full dynamic theory
in our earlier work\cite{2015arXiv151005510G} we considered
a probability state in time space $\psi(t)$
with two measures(Laguerre and Shifted Legendre)
form Section II ``Kinematics'' of Ref.  \cite{2015arXiv151005510G}
having extremely convenient properties that allows to obtain
infinitesimal time--shift variation by applying integration by parts.
The question arise about combining the $\psi(t)$ and $\psi(p)$
probability states into a single state, that would
describe price dynamics, and give an answer of 
how much information about future price values can be actually obtain,
i.e. to what extend a trader dream of ``philosopher stone'', the system that predict future prices, can actually exist.
This would be the topic of our future research.

\section{\label{discuss} Discussion}
In this paper an alternative to Supply--Demand theory
is proposed. The theory is considering
always matched buyers and sellers,
and maximizing the rate of their matching.
System state is determined by 
a probability distribution,
from which  all observable variable,
including matching rate, can be calculated.
The equilibrium distribution, corresponding to maximal matching rate,
can be found from generalized eigenvalues problem,
maximizing the matching rate functional (\ref{IpsiG}).
An application of the theory is demonstrated on AAPL
intraday trading data. A conceptual difference
between maximizing the trading volume
and matching rate (trading volume per unit time)
is shown. While the trading volume has maximal values about median price,
the matching rate has a singularity--like behavior near the market
tipping points, what make the approach much more suitable
to risk measurement and market direction prediction.

\appendix
\section{\label{code}Code implementation example}
Computer code implementing the algorithms is available\cite{polynomialcode}.
The code is java/scala written. To reproduce the results
follow these steps:
\begin{itemize}
\item Install java 1.8 or later \& scala  2.11.7 or later.
\item Download from \cite{polynomialcode}  the data file S092012-v41.txt.gz
  and code archive SupplyDemandQuasiStationary.zip.
\item Decompress the code and recompile it.
\begin{verbatim}
    unzip SupplyDemandQuasiStationary.zip
    javac -g com/polytechnik/*/*java
    scalac com/polytechnik/algorithms/ExampleNoSupplyDemand.scala
\end{verbatim}
\item Extract from the file S092012-v41.txt.gz AAPL executed trades
  and save the data to aapp.csv
\begin{verbatim}
java com/polytechnik/itch/DumpDataTrader S092012-v41.txt.gz AAPL >aapl.csv
\end{verbatim}
The $P(t)$ in Fig. \ref{fig:P} can be obtained from the aapl.csv,
the unit of time (second column) in aapl.csv is the
one used in NASDAQ ITCH\cite{itchfeed}: nanoseconds since 00:00.
\item Use executed trades information from aapl.csv to obtain
 Section \ref{numestim} results.
\begin{verbatim}
    scala com.polytechnik.algorithms.ExampleNoSupplyDemand aapl.csv
\end{verbatim}
The files histogram.csv and psi.csv are now generated.
The file histogram.csv contains price--volume distribution
(top chart in Fig. \ref{fig:EQl}).
The file psi.csv contains $w^{[H]}(P)$, $w^{[L]}(P)$ and $I(P)$
(other charts in Fig. \ref{fig:EQl}).
\end{itemize}

\bibliography{LD}

\end{document}